\documentclass[12pt]{article}
\usepackage{graphicx}
\def\ba{\begin{eqnarray}}
\def\ea{\end{eqnarray}}
\def\bas{\begin{eqnarray*}}
\def\eas{\end{eqnarray*}}
\def\be{\begin{equation}}
\def\ee{\end{equation}}

\def\nnr{\nonumber \\}

\catcode`@=12
\begin{document}
\thispagestyle{empty}
\begin{flushright}
\end{flushright}
\vspace{0.5in}
\begin{center}
{\LARGE \bf Quark and lepton mass matrices\\[0.2in] with $A_4$ family symmetry}
\footnote{Talk given at  International Workshop on Neutrino Masses and Mixings, University of Shizuoka, Shizuoka, Japan, 17-19 December 2006. }\\
\vspace{1.5in}
{\bf Hideyuki Sawanaka}
\footnote{e-mail address: hide@muse.sc.niigata-u.ac.jp}\\
\vspace{0.2in}
{\sl Graduate School of Science and Technology, Niigata University,
Ikarashi 2-8050, Niigata 950-2181, Japan\\}
\vspace{1.5in}
\end{center}

\begin{abstract}\
Realistic quark masses and mixing angles are obtained applying
the successful $A_4$ family symmetry for leptons, 
motivated by the quark-lepton assignments of SU(5).
The $A_4$ symmetry is suitable to give tri-bimaximal neutrino mixing matrix 
which is consistent with current experimental data.  
We study new scenario for the quark sector with the $A_4$ symmetry.  
\end{abstract}

\clearpage
\baselineskip 24pt

\section{Introduction}

The current observed neutrino mixing \cite{ex1,ex2} suggests around the 
maximal 2-3 mixing angle and zero 1-3 mixing angle: 
$\theta_{23} \sim \pi/4$, $\theta_{13} \sim 0$.  
In such a symmetric limit where both $\cos\theta_{23}$ and 
$|U_{e3}|=\sin\theta_{13}$ vanish, the resulting $3 \times 3$ effective 
Majorana neutrino mass matrix forms in the flavor basis as \cite{GJKLST} 
\be
\left(\begin{array}{ccc} 
 X & C & C \\ 
 C & A & B \\ 
 C & B & A 
\end{array}\right).  
\label{M-Z2}
\ee
This matrix has an exact symmetric form under a $Z_2$ symmetry, i.e. 
the 2-3 ($\mu$-$\tau$) permutation, 
and  is diagonalized by the unitary matrix: 
\be
 U_{Z_2} = 
 \left(\begin{array}{ccc} 
 \cos\theta_{12} & -\sin\theta_{12} & 0 \\ 
 \sin\theta_{12}/\sqrt{2} & \cos\theta_{12}/\sqrt{2} & -1/\sqrt{2} \\ 
 \sin\theta_{12}/\sqrt{2} & \cos\theta_{12}/\sqrt{2} & 1/\sqrt{2} 
 \end{array}\right), 
\ee
with remaining the solar mixing angle $\theta_{12}$ arbitrary 
and the entry $X$ of (\ref{M-Z2}) is determined as
\be
 X = A + B + \frac{2\sqrt{2}C}{\tan 2\theta_{12}}.  
\ee
Now we well know  two special values for $\theta_{12}$ 
which give typical mixing matrices; 
one is bimaximal and the other is tri-bimaximal mixing matrices.  
In a limit of bimaximal mixing \cite{Barger:1998ta} 
where $\theta_{12} = 4/\pi$, resulting MNS matrix forms 
\be
 U_{\rm BM} =  \left(\begin{array}{ccc} 
 1/\sqrt{2} & -1/\sqrt{2} & 0 \\ 
 1/2 & 1/2 & -1/\sqrt{2} \\ 
 1/2 & 1/2 & 1/\sqrt{2} \end{array}\right), 
\ee
with 
\be
 X = A + B.  
\ee
For the case $\tan\theta_{12} = 1/\sqrt{2}$ with 
so-called tri-bimaximal mixing which is proposed by
Harrison, Perkins and Scott,  
then we have the HPS type matrix \cite{HPS,Xing:2002sw}: 
\be
U_{\rm HPS} = \left(\begin{array}{ccc} 
 2/\sqrt{6}   &  1/\sqrt{3}  &  0  \\ 
 -1/\sqrt{6}  &  1/\sqrt{3}  & -1/\sqrt{2} \\ 
 -1/\sqrt{6}  &  1/\sqrt{3}  &  1/\sqrt{2} 
\end{array}\right), 
\ee
where $\theta_{12}$ is fixed by as well 
\be
 X = A + B + C
\ee
is also derived.  
Note that the tri-bimaximal structure is consistent with current experimental data, where $\theta_{12}$ is not maximal.  

So far, the discrete symmetry $A_4$ is successfully applied for
leptons. Namely, the tri-bimaximal mixing pattern can be realized
naturally in a number of specific models \cite{A4-1,A4-2,A4-3,A4-4,A4-5}.  
However, it is not easy to have small quark mixing angles
in naive and straight way.  In such applications the generic
prediction \cite{A4-2} is that $V_{\rm CKM}$, 
the quark mixing matrix becomes just the unit matrix.  
Starting with $V_{\rm CKM} = {\bf 1}$, the realistic small
quark mixing angles can be generated 
by extending interactions beyond those of the Standard Model, such as in
supersymmetry \cite{A4-3} or breaking $A_4$ symmetry explicitly \cite{A4-4}.   
Our study is aimed to obtain realistic quark masses and mixing angles 
entirely within the $A_4$ context \cite{MST}.  
It is worthy of mention that 
the other types of unified models for quarks and leptons 
with the $A_4$ symmetry \cite{gut-Ma,gut-King} and 
models which predict tri-bimaximal mixing matrix 
with $S_3$ symmetry \cite{S3-GL,S3-HWY,S3-Mohapatra} have been also studied.

\section{$A_4$ symmetry}

$A_4$ is the symmetry group of the tetrahedron and 
the finite groups of the even permutation of four objects.  
It has twelve elements which are derived into four equivalence classes: 
[$C_1$]: (1234), 
[$C_2$]: (2143), (3412), (4321), 
[$C_3$]: (1342), (4213), (2431), (3124) and 
[$C_4$]: (1423), (3241), (4132), (2314),
corresponding to its four irreducible representations
we call three one-dimensional representations (singlets) 
as $\underline{1}$, $\underline{1}^\prime$, 
$\underline{1}^{\prime\prime}$, and 
one three-dimensional representation (triplet) as $\underline{3}$, 
respectively.  
The $A_4$ is the smallest discrete group 
which includes the three-dimensional irreducible representation.  
The presence of the three-dimensional irreducible representation 
might be ideal for describing three families of quarks and leptons.  
The character table of four representations is 
shown in Table~\ref{table:ct}.  
Here $h$ is the order of each element, $n$ is the number of elements and 
the complex number $\omega$ is the cube root of unity: 
\be
 \omega = \exp(2 \pi i/3) = - \frac{1}{2} + \frac{\sqrt{3}}{2} i, \quad 
 1 + \omega + \omega^2 = 0.  
\ee
\begin{table}[pt]
\begin{center}
\begin{tabular}{@{}ccccccc@{}} \hline 
 &$h$&$n$&$\underline{1}$&$\underline{1}^\prime$&$\underline{1}^{\prime\prime}$&$\underline{3}$ \\ \hline
  $C_1$&1&1&1&1&1&3 \\ 
  $C_2$&2&3&1&1&1&$-$1 \\ 
  $C_3$&3&4&1&$\omega$&$\omega^2$&0 \\   
  $C_4$&3&4&1&$\omega^2$&$\omega$&0 \\ \hline
\end{tabular}
\caption{Character table of $A_4$.  \label{table:ct}}
\end{center}
\end{table}

The fundamental multiplication rules are given as%
\footnote{For details of the $A_4$ multiplication rules, see the original paper \cite{A4-1,A4-2} for example.  }
\be
 \underline{1}^\prime \times \underline{1}^\prime =
 \underline{1}^{\prime\prime}, \quad 
 \underline{1}^{\prime\prime} \times \underline{1}^{\prime\prime} = 
 \underline{1}^\prime, \quad 
 \underline{1}^\prime \times \underline{1}^{\prime\prime} = 
 \underline{1}, 
\label{deco1x1}
\ee
and 
\be
 \underline{3}_1 \times \underline{3}_2 = 
 \underline{1} + \underline{1}^\prime +\underline{1}^{\prime\prime} + 
 \underline{3}_{\rm A} + \underline{3}_{\rm B}, 
\label{deco3x3}
\ee
where denoting $\underline{3}_i$ for $i=1$, $2$ as $(a_i, b_i, c_i)$, 
we have 
\bas
 \underline{1} &\sim& a_1a_2 + b_1b_2 + c_1c_2, \\
 \underline{1}^\prime &\sim& a_1a_2 + \omega b_1b_2 + \omega^2 c_1c_2, \\
 \underline{1}^{\prime\prime} &\sim& a_1a_2 + \omega^2 b_1b_2 + \omega c_1c_2, \\
 \underline{3}_{\rm A} &\sim& (b_1c_2, c_1a_2, a_1b_2), \\ 
 \underline{3}_{\rm B} &\sim& (c_1b_2, a_1c_2, b_1a_2).  
\eas
Note that from Eq.~(\ref{deco1x1}) the $A_4$ invariant singlet 
$\underline{1}$ can be derived in these three sets of $A_4$ singlets: 
\be
 \underline{1}^\prime \times \underline{1}^\prime \times \underline{1}^\prime, \quad 
 \underline{1}^{\prime\prime} \times \underline{1}^{\prime\prime} \times \underline{1}^{\prime\prime}, \quad  
 \underline{1} \times \underline{1}^\prime \times \underline{1}^{\prime\prime}, 
\label{1x1x1}
\ee
and 
$\underline{3} \times \underline{3} \times \underline{3} = \underline{1}$ 
is also possible in the $A_4$ symmetry from Eq.~(\ref{deco3x3}).  
By using them, $A_4$ invariant mass matrices are constructed.

\section{$A_4$ model for lepton sector}

In this section, we briefly show a simple example of the $A_4$ model 
for leptons and lead the tri-bimaximal mixing matrix.  For 
more detail of advanced models can be found 
in recent reviews \cite{Mohapatra:2006gs,Altarelli:2006ri}.  

Let us take the $A_4$ assignment for leptons as shown in Table~\ref{table:asl}: 
left-handed SU(2)$_L$ lepton doublets $(\nu_i, l_i)$ ($i=1, 2, 3$) 
transform as an $A_4$ triplet $\underline{3}$, 
while right-handed, charged lepton singlets $l_i^c$ transform as $A_4$ singlets
($l_1^c=e^c$, $l_2^c=\mu^c$ and $l_3^c=\tau^c$ transform as 
$\underline{1}$, $\underline{1}^\prime$ and 
$\underline{1}^{\prime\prime}$, respectively).  
\begin{table}[pt]
\begin{center}
\begin{tabular}{@{}cccccc@{}} \hline
 &$(\nu_i, l_i)$&$e^c$&$\mu^c$&$\tau^c$&
  $\phi_{li}$\\ \hline
  $A_4$ &$\underline{3}$&$\underline{1}$&$\underline{1}^\prime$&$\underline{1}^{\prime\prime}$&$\underline{3}$ \\ \hline  
\end{tabular}
\caption{$A_4$ assignment for leptons.  \label{table:asl}}
\end{center}
\end{table}
Introducing gauge singlet Higgs doublet 
$\phi_{li} = (\phi_{li}^0, \phi_{li}^-) \sim \underline{3}$ 
under $A_4$, the $3 \times 3$ mass matrix 
linking $l_i$ with $l_i^c$ is given by 
\be
M_l = \left(\begin{array}{ccc}
 f_1v_{l1} & f_2v_{l1} & f_3v_{l1} \\ 
 f_1v_{l2} & f_2\omega v_{l2} & f_3\omega^2 v_{l2} \\ 
 f_1v_{l3} & f_2\omega^2 v_{l3} & f_3\omega v_{l3}
\end{array}\right) 
= 
\left(\begin{array}{ccc}
 v_{l1} & 0 & 0 \\ 
 0 & v_{l2} & 0 \\ 
 0 & 0 & v_{l3}
\end{array}\right)
\left(\begin{array}{ccc}
 1 & 1 & 1 \\ 
 1 & \omega & \omega^2 \\ 
 1 & \omega^2 &\omega
\end{array}\right)
\left(\begin{array}{ccc}
 f_1 & 0 & 0 \\ 
 0 & f_2 & 0 \\ 
 0 & 0 & f_3
\end{array}\right), 
\ee
where $\omega = \exp(2 \pi i/3)$, $f_i$ are Yukawa couplings 
and $v_{li} = \langle \phi_{li}^0 \rangle$ are 
vacuum expectation values of the Higgs field $\phi_{li}^0$.  It can 
be diagonalized in a very simple way, i.e. by setting all 
three vacuum expectation values of the Higgs field to be equal.  Taking 
$v_l \equiv v_{l1} = v_{l2} = v_{l3}$, 
the charged lepton mass matrix is diagonalized as 
\be
M_l^{diag} = 
U_{lL}^\dag M_l U_{lR}= \left(\begin{array}{ccc}
 f_1 & 0 & 0 \\ 
 0 & f_2 & 0 \\ 
 0 & 0 & f_3
\end{array}\right)
\sqrt{3}v_l = \left(\begin{array}{ccc}
 m_e & 0 & 0 \\ 
 0 & m_\mu & 0 \\ 
 0 & 0 & m_\tau
\end{array}\right), 
\label{Ml-diag}
\ee
where
\be
U_{lL} = \frac{1}{\sqrt{3}}\left(\begin{array}{ccc}
 1 & 1 & 1 \\ 
 1 & \omega & \omega^2 \\ 
 1 & \omega^2 &\omega
\end{array}\right), 
\ee
and $U_{lR}$ is the unit matrix.  
Three different charged lepton masses are given by the Yukawa couplings.

For the neutrino mass matrix, if we consider 
six gauge singlet Higgs triplets 
$\xi_i = (\xi_i^{++}, \xi_i^+, \xi_i^0)$ which are assigned to 
$\xi_1 \sim \underline{1}$, $\xi_2 \sim \underline{1}^\prime$, 
$\xi_3 \sim \underline{1}^{\prime\prime}$ and 
$\xi_{4,5,6} \sim \underline{3}$ under $A_4$, 
then the matrix forms in general 
\be
M_\nu = \left(\begin{array}{ccc}
 a+b+c & f & e \\ 
 f & a+\omega b+\omega^2 c & d \\ 
 e & d & a+\omega^2 b+\omega c
\end{array}\right), 
\ee
here parameters $(a, b, c)$ come from 
$(\xi_1, \xi_2, \xi_3)$ and $(d, e, f)$ from $\xi_{4,5,6}$, respectively.  
If conditions $b=c$ and $e=f=0$ are given 
(hopefully in some natural mechanisms), 
$M_\nu$ is diagonalized by the matrix $U_{\nu L}$ 
with three eigenvalues (neutrino masses) 
$m_1 = a-b+d$, $m_2 = a+2b$, $m_3 = -a+b+d$, and then we have 
\be
U_{\rm MNS} = (U_{lL})^\dag U_{\nu L} = \left(\begin{array}{ccc} 
 2/\sqrt{6}   &  1/\sqrt{3}  &  0  \\ 
 -1/\sqrt{6}  &  1/\sqrt{3}  & -1/\sqrt{2} \\ 
 -1/\sqrt{6}  &  1/\sqrt{3}  &  1/\sqrt{2} 
\end{array}\right).  
\ee
which is exactly the tri-bimaximal mixing matrix.

\section{$A_4$ model for quark sector}

Following the successful lepton assignments in the previous section, 
if we assign for quarks as in Table~\ref{table:a4sq} 
and take $v_{q1} = v_{q2} = v_{q3} \equiv v_q$
($v_{qi} = \langle \phi_{qi}^0 \rangle$), 
we then have the up and down quark mass matrices as 
\be
M^{U(D)} = \frac{1}{\sqrt{3}}\left(\begin{array}{ccc}
 1 & 1 & 1 \\ 
 1 & \omega & \omega^2 \\ 
 1 & \omega^2 &\omega
\end{array}\right)
\left(\begin{array}{ccc}
 g_1^{U(D)} & 0 & 0 \\ 
 0 & g_2^{U(D)} & 0 \\ 
 0 & 0 & g_3^{U(D)}
\end{array}\right)
\sqrt{3}v_q.  
\ee
\begin{table}[pt]
\begin{center}
\begin{tabular}{@{}cccc@{}} \hline
 &$(u_i, d_i)$&$u_i^c$, $d_i^c$&$(\phi_{qi}^0, \phi_{qi}^-)$ \\ \hline
  $A_4$ &$\underline{3}$&$\underline{1}$, $\underline{1}^\prime$, $\underline{1}^{\prime\prime}$&$\underline{3}$ \\ \hline  
\end{tabular}
\caption{$A_4$ assignment for quarks which follows that for leptons.  \label{table:a4sq}}
\end{center}
\end{table}
The generic prediction is that 
the quark mixing matrix leads to the unit matrix: 
$V_{\rm CKM} = (U^U)^\dag U^D = {\bf 1}$ where
\be
U^U = U^D = \frac{1}{\sqrt{3}}\left(\begin{array}{ccc} 
 1 & 1 & 1 \\ 
 1 & \omega & \omega^2 \\ 
 1 & \omega^2 &\omega 
\end{array}\right) 
\ee
are matrices which diagonalize $M^{U(D)}$ as (\ref{Ml-diag}).  
The realistic small quark mixing angles can be generated 
by extending interactions beyond those of the Standard Model, 
such as in supersymmetry \cite{A4-4} or the addition of terms which 
break the $A_4$ symmetry as well as the residual $Z_3$ symmetry 
explicitly \cite{A4-5}.  In this section, we present 
a new alternative scenario, 
where realistic quark masses and mixing angles are obtained
motivated by the quark-lepton assignments of SU(5), 
entirely within the $A_4$ context \cite{MST}.  

In SU(5) grand unification, the $\underline{5}^\ast$ representation 
contains the lepton doublet $(\nu,l)$ and the quark singlet $d^c$, 
whereas the $\underline{10}$ representation contains 
the lepton singlet $l^c$ and the quark doublet $(u,d)$ and singlet $u^c$.  
In the successful $A_4$ model for leptons, 
$(\nu_i,l_i)$ transform as $\underline{3}$ whereas 
$l^c_i$ transform as $\underline{1}$, $\underline{1}^\prime$ 
and $\underline{1}^{\prime\prime}$.  
Thus we choose as shown in Table~\ref{table:assu5}.  
\begin{table}[pt]
\begin{center}
\begin{tabular}{@{}ccccc@{}} \hline
  &$(\nu_i, l_i), d^c_i$&$l^c_i, u^c_i, (u_i, d_i)$&$(\phi^0_{U1,2}, \phi^-_{U1,2})$
  &$(\phi^0_{li}, \phi^-_{li}), (\phi^0_{Di}, \phi^-_{Di})$ 
  \\ \hline
 $A_4$ &$\underline{3}$&$\underline{1}$, $\underline{1}^\prime$, $\underline{1}^{\prime\prime}$&$\underline{1}^\prime$, $\underline{1}^{\prime\prime}$&$\underline{3}$ \\ 
   SU(5)&$\underline{5}^\ast$&$\underline{10}$&$\underline{5}$&$\underline{5}^\ast + \underline{45}$ \\ \hline 
\end{tabular}
\caption{SU(5) motivated quark and lepton assignment.  \label{table:assu5}}
\end{center}
\end{table}
In minimal SU(5), there is just one $\underline{5}$ representation of 
Higgs bosons, yielding thus only two invariants, i.e. 
$\underline{10} \times \underline{10} \times \underline{5} \to \underline{1}$ 
(that is for up quark mass matrix) and 
$\underline{5}^\ast \times \underline{10} \times \underline{5}^\ast 
\to \underline{1}$ 
(for charged lepton and down quark mass matrices).  The 
second invariant implies $m_\tau = m_b$ at the unification scale which is 
phenomenologically desirable.  
We follow the usual strategy of 
using both $\underline{5}^\ast$ and $\underline{45}$ 
representations of Higgs bosons, so that one linear combination 
couples to only leptons, and the other only to quarks.  
Both transform as $\underline{3}$ under $A_4$.  
There are also two $\underline{5}$ representations transforming as 
$\underline{1}^\prime$ and $\underline{1}^{\prime\prime}$ under $A_4$ 
which couple only to up type quarks%
\footnote{We choose here two Higgs doublets transforming as $\underline{1}^\prime$ and $\underline{1}^{\prime\prime}$, other patterns can be assigned with different prediction.  }.  

With this $A_4$ assignment for Higgs doublets, 
the relevant Yukawa couplings linking $d_i$ with $d^c_j$ are given by
\bas
&&h_1 d_1 (d^c_1 \phi^0_{D1} + d^c_2 \phi^0_{D2} + d^c_3 \phi^0_{D3}) + 
h_2 d_2 (d^c_1 \phi^0_{D1} + \omega d^c_2 \phi^0_{D2} + \omega^2 d^c_3 \phi^0_{D3}) \\ 
&& +~h_3 d_3 (d^c_1 \phi^0_{D1} + \omega^2 d^c_2\phi^0_{D2} + \omega d^c_3 \phi^0_{D3}),
\eas
\noindent resulting in the $3 \times 3$ down quark mass matrix:
\be
M_{\rm D} = \left(\begin{array}{ccc}
 {h_1} & 0 & 0 \\ 
 0 & {h_2} & 0 \\ 
 0 & 0 & {h_3}
\end{array}\right)
\left(\begin{array}{ccc}
 1 & 1 & 1 \\ 
 1 & \omega & \omega^2 \\ 
 1 & \omega^2 &\omega
\end{array}\right)
\left(\begin{array}{ccc}
 {v_1} & 0 & 0 \\ 
 0 & {v_2} & 0 \\ 
 0 & 0 & {v_3}
\end{array}\right), 
\ee
where $\omega = \exp(2 \pi i/3)$, $h_i$ are three independent Yukawa 
couplings, and $v_i$ are the vacuum expectation values of $\phi_{{\rm D}i}^0$.  
To get quark mass hierarchy, 
$v_1 \ll v_2 \ll v_3$ should be satisfied contrary to the lepton sector
($v_{li} \equiv \langle \phi^0_{li} \rangle = v_l$).  
By contrast, the Higgs doublets transforming as $\underline{1}^\prime$ and 
$\underline{1}^{\prime\prime}$ linking $u_i$ with $u_j^c$ give, 
as following the manner of Eq.~(\ref{1x1x1}), 
the $3 \times 3$ symmetric up quark mass matrix:
\be
M_{\rm U} = \left(\begin{array}{ccc} 
 0 & {\mu_2} & {\mu_3} \\ 
 {\mu_2} & {m_2} & 0 \\ 
 {\mu_3} & 0 & {m_3} 
\end{array}\right), 
\ee
where $m_2$, $\mu_3$ come from 
$\phi_{{\rm U}1}^0 \sim \underline{1}^\prime$ and $m_3$, $\mu_2$ from 
$\phi_{{\rm U}2}^0 \sim \underline{1}^{\prime\prime}$ 
and three of them can be taken real in general.  
In the limit $|\mu_2| \ll |m_2|$ and $|\mu_3| \ll |m_3|$, we obtain the 
three eigenvalues of $M_{\rm U}$ as 
\be
m_t \simeq |m_3|, \quad 
m_c \simeq |m_2|, \quad 
m_u \simeq \left|\frac{\mu_2^2}{m_2}+\frac{\mu_3^2}{m_3} \right|, \quad 
\ee
with mixing angles 
\be
V_{uc} \simeq \frac{\mu_2}{m_2}, \quad 
V_{ut} \simeq \frac{\mu_3}{m_3}, \quad 
V_{ct} \simeq 0.  
\ee

In the down sector, we note that 
\be
M_{\rm D}M_{\rm D}^\dag = \left(\begin{array}{ccc}
Y |h_1|^2 & Z^\ast h_1 h_2^\ast & Z h_1 h_3^\ast \\ 
Z h_1^\ast h_2 & Y |h_2|^2 & Z^\ast h_2 h_3^\ast \\ 
Z^\ast h_1^\ast h_3 & Z h_2^\ast h_3 & Y |h_3|^2
\end{array}\right),
\ee
where 
\bas
 Y &=& |v_1|^2 + |v_2|^2 + |v_3|^2, \\
 Z &=& |v_1|^2 + \omega |v_2|^2 + \omega^2 |v_3|^2.  
\eas
Its eigenvalue $\lambda$ satisfies the equation
\ba
&&\lambda^3 - Y(|h_1|^2 + |h_2|^2 + |h_3|^2)\lambda^2 
 - (Y^3+Z^3+Z^{\ast 3} -3Y|Z|^2)|h_1|^2|h_2|^2|h_3|^2 \nnr
&& +~(Y-|Z|^2)(|h_1|^2|h_2|^2 + |h_1|^2|h_3|^2 + |h_2|^2|h_3|^2)\lambda 
= 0.  
\ea
If $|v_1|=|v_2|=|v_3|=|v|$ as well as assumed in the charged lepton case, 
then $Y=3|v|^2$, $Z=0$ and three eigenvalues are simply 
$3|h_{1, 2, 3}|^2|v|^2$.  We choose them instead to be different, 
but we still assume $|h_1|^2 \ll |h_2|^2 \ll |h_3|^2$.  In that case, 
we find 
\ba
 m_b^2 &\simeq& Y|h_3|^2, \\
 m_s^2 &\simeq& \left(\frac{Y^2-|Z|^2}{Y}\right)|h_2|^2, \\
 m_d^2 &\simeq& \left(\frac{Y^3+Z^3+Z^{\ast 3}-3Y|Z|^2}{Y^2-|Z|^2}\right)|h_1|^2, 
\ea
and the mixing angles are given by
\ba
 V_{sb} &\simeq& \left( \frac{Z^\ast}{Y} \right) \frac{h_2}{h_3}, \\ 
 V_{db} &\simeq& \left( \frac{Z}{Y} \right) \frac{h_1}{h_3}, \\ 
 V_{ds} &\simeq& \left( \frac{YZ^\ast - Z^2}{Y^2-|Z|^2} \right) \frac{h_1}{h_2},
\ea
thereby requiring the condition
\be
 \left| \frac{V_{ds}V_{sb}}{V_{db}} \right| \simeq 
 \left| \frac{Y Z^* - Z^2}{Y^2 - |Z|^2} \right|.  
\ee
Using current experimental values for the left-hand side, we see that 
quark mixing in the down sector alone cannot explain the observed 
quark mixing matrix $V_{\rm CKM}$.  Taking into account $V_{\rm U}$, 
we then have 
\be
 V_{\rm CKM}=V_{\rm U}^\dag V_{\rm D}.  
\ee
Hence 
\ba
 V_{us} &\simeq& V_{ds} - V_{uc} \simeq 
 \left( \frac{YZ^\ast - Z^2}{Y^2-|Z|^2} \right) \frac{h_1}{h_2} 
 - \frac{\mu_2}{m_2}, 
\label{Vus} \\ 
 V_{cb} &\simeq& V_{sb} \simeq 
 \left( \frac{Z^\ast}{Y} \right) \frac{h_2}{h_3}, 
\label{Vcb} \\ 
 V_{ub} &\simeq& V_{db} - V_{uc} V_{sb} - V_{ut} \simeq 
 \left( \frac{Z}{Y} \right) \frac{h_1}{h_3} 
 - \left( \frac{Z^\ast}{Y} \right) \frac{h_2}{h_3} \frac{\mu_2}{m_2} 
 - \frac{\mu_3}{m_3}.  
\label{Vub}
\ea

It is noted that 
our up and down quark mass matrices are restricted 
by our choice of $A_4$ representations 
to have only five independent parameters each.  
In the up sector, we have three real and one complex parameters 
(for example we choose here 
$m_2$, $m_3$ and $\mu_2$ to be real with $\mu_3$ complex).  
The five independent parameters can be chosen as 
the three up quark masses, one mixing angle and one phase.   
In the down sector, 
the Yukawa couplings $h_{1, 2, 3}$ can all be chosen real, 
$Y$ is just an overall scale, and $Z$ is complex.  
The five independent parameters can be chosen as 
the three down quark masses and two mixing angles.  
Now we have ten parameters in these two matrices 
except their overall normalizations or magnitudes of Yukawa couplings.  
Since we also have ten observables for six quark masses, three angles 
and one phase, it may appear that a fit is not so remarkable.  
However, the forms of the mass matrices are very restrictive, 
and it is by no means trivial to obtain a good fit.  
Indeed, we find that $V_{ub}$ is strongly correlated with the CP phase $\beta$ which is one of  angles of the unitary triangle.  
If we were to fit just the six masses and the three angles, 
the structure of our mass matrices would allow only a very narrow range 
of values for $\beta$ at each value of $|V_{ub}|$.  
This means that future more precise determinations of 
these two parameters will be a decisive test of this model.

CP violation is also predicted in our model.  
The Jarlskog invariant \cite{Jarlskog:1985ht} is given by 
\be
 J_{\rm CP}\simeq \frac{\sqrt{3}}{2}\frac{h_1^2}{h_3^2} 
 \left( \frac{v_2^2 - v_1^2}{v_2^2 + v_1^2} \right) 
 \left (1+ \frac{{\rm Re}(\mu_3)+\frac{1}{\sqrt{3}}{\rm Im}(\mu_3)}{m_t} 
 \frac{h_3}{h_1} \right ), 
\label{Jcp}
\ee
which is mainly comes from the down sector and 
the up sector only affects as correction terms.  

\begin{figure}[th]
\centerline{
\includegraphics{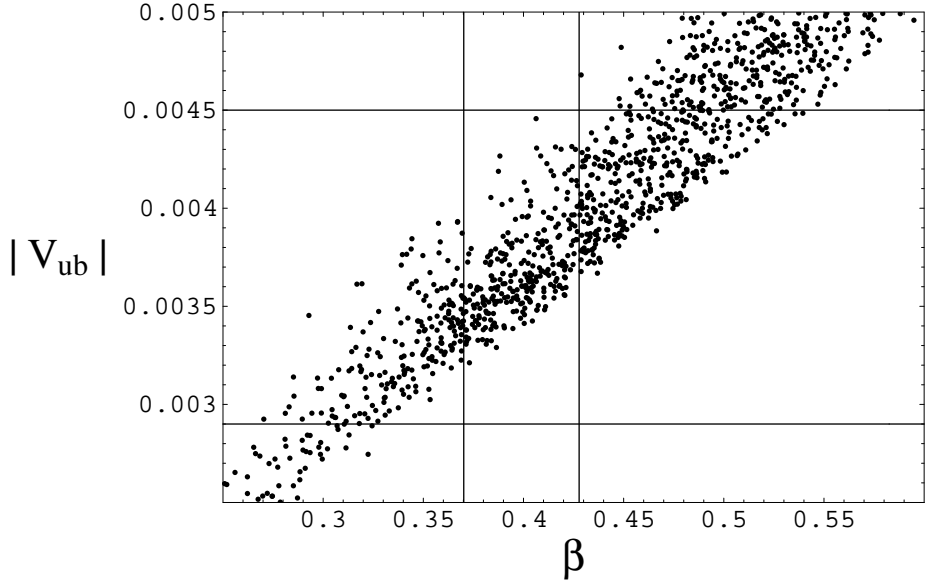}
}
\vspace*{8pt}
\caption{Plot of allowed values  in the $\beta - |V_{ub}|$ plane, where the value of $\beta$ is expressed in radians.  The horizontal and vertical lines denote experimental bounds at $90\%$ C.L.~(\ref{input1}).  }
\label{f1}
\end{figure}

\begin{figure}[th]
\centerline{\includegraphics{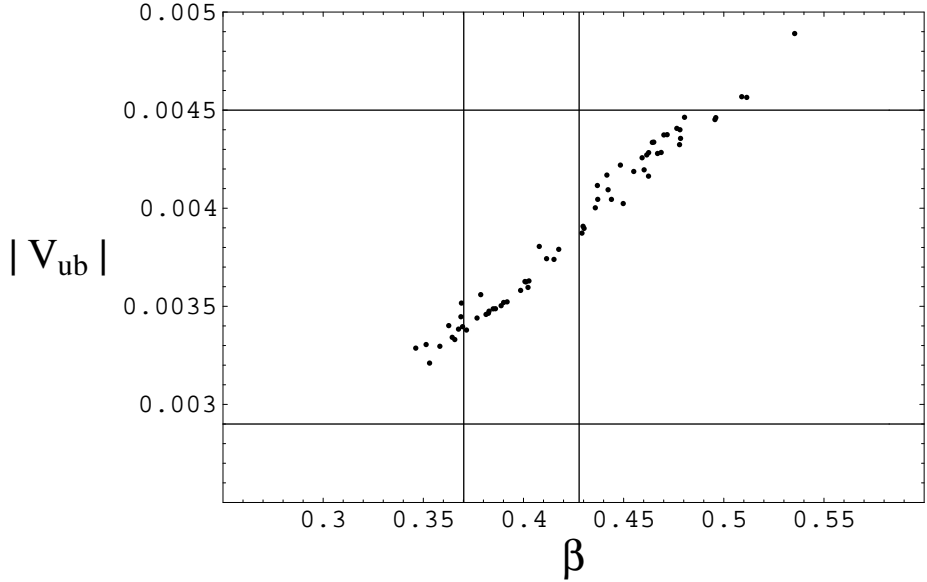}}
\vspace*{8pt}
\caption{Plot of allowed values  in the $\beta - |V_{ub}|$ plane, where input data are restricted in the narrower regions shown in Eq~(\ref{input2}).  }
\label{f2}
\end{figure}

In order to fit the ten observables 
(six quark masses, three CKM mixing angles and one CP phase), 
ten parameters of our model have been generated numerically.  
We choose the parameter sets which are allowed by the experimental data.  
We show the prediction of $|V_{ub}|$ verses $\beta$ in Fig.~\ref{f1}, 
with the following nine experimental inputs 
\cite{Fritzsch:1999ee,Yao:2006px,Charles:2006yw}:  
\ba
&& 
m_u = 0.9 \sim 2.9 \ {\rm (MeV)},\quad 
m_c = 530 \sim 680 \ {\rm (MeV)}, \quad 
m_t = 168 \sim 180 \ {\rm (GeV)}, 
\nnr && 
m_d = 1.8 \sim 5.3\ {\rm (MeV)}, \quad 
m_s = 35 \sim 100 \ {\rm (MeV)}, \quad 
m_b = 2.8 \sim 3 \ {\rm (GeV)}, 
\nnr && 
|V_{us}| = 0.221 \sim 0.227, \quad 
|V_{cb}| = 0.039 \sim 0.044, \quad 
J_{\rm CP} = (2.75 \sim 3.35)\times 10^{-5}, 
\nnr && 
\label{input1}
\ea
which are given at the electroweak scale.  
We see that the experimental allowed region of $\beta$ 
($0.370 \sim 0.427$ radian  at $90\%$ C.L.) \cite{Charles:2006yw} 
corresponds to $|V_{ub}|$ in the range $0.0032\sim 0.0044$, 
which is consistent with the experimental value of 
$|V_{ub}| = 0.0029 \sim 0.0045$.  Thus our model is able to reproduce 
realistically the experimental data of quark masses and the CKM matrix.
\indent
Precisely measured heavy quark masses and CKM matrix elements 
are expected in future experiments and precise light quark masses 
are expected in future lattice evaluations.   If the allowed regions 
of the current data shown in Eq.~(\ref{input1}) are reduced,
the correlation between  $|V_{ub}|$ and $\beta$ will become stronger. 
We show in Fig.~\ref{f2} the case where the experimental data are 
restricted to some very narrow ranges about their central values: 
\ba
&& 
 m_u = 1.4 \sim 1.5 \ {\rm (MeV)}, \quad 
 m_c = 600 \sim 610 \ {\rm (MeV)}, \quad 
 m_t = 172 \sim 176 \ {\rm (GeV)}, 
\nnr && 
 m_d = 3.4 \sim 3.6\ {\rm (MeV)}, \quad 
 m_s = 60 \sim 70 \ {\rm (MeV)}, \quad 
 m_b = 2.85 \sim 2.95 \ {\rm (GeV)}, 
\nnr && 
 |V_{us}| = 0.221 \sim 0.227, \quad 
 |V_{cb}| = 0.041 \sim 0.042, \quad 
 J_{\rm CP} = (3.0 \sim 3.1) \times 10^{-5}, 
\nnr && 
\label{input2}
\ea
Here we use the tighter constraints on the mass ratios of light quarks, 
i.e. $m_u/m_d$ and $m_s/m_d$, consistent with the well-known successful 
low-energy sum rules \cite{Leutwyler:1996qg}:  
Clearly, future more precise determinations of 
$|V_{ub}|$ and $\beta$ will be a sensitive test of our model.  

A comment is in order.
Our quark mass matrices are in principle given at 
the SU(5) unification scale.  However, the $A_4$ flavor symmetry is 
spontaneously broken at the electroweak scale.  Therefore, 
the forms of our mass matrices are not changed except for 
the magnitudes of the Yukawa couplings between the unification and 
electroweak scales.  Hence, our numerical analyses are presented 
at the electroweak scale.

We should also comment on the hierarchy of $h_i$ and $v_i$.  
The order of $h_i$ are fixed by the quark mixing (\ref{Vus}), (\ref{Vcb}).
The ratios of 
\be
h_1/h_2 \simeq \lambda (\simeq 0.22), \quad 
h_2/h_3 \simeq \lambda^2, 
\ee
are required by  $V_{us}$ and $V_{cb}$, respectively.  
Once $h_i$ are fixed, quark masses determine the hierarchy of $v_i$ 
as follows:
\be
v_1/v_3 \simeq \lambda^2, \quad 
v_2/v_3 \simeq \lambda \sim \lambda^{1/2}.  
\ee
These hierarchies of $h_i$ and $v_i$ are also consistent with 
the magnitude of $J_{\rm CP}$ given in Eq.~(\ref{Jcp}).

\section{Summary}

The $A_4$ family symmetry which has been successful in understanding 
the mixing pattern of neutrinos (tri-bimaximal mixing) 
is applied to quarks, 
motivated by the quark-lepton assignments of SU(5).  Realistic quark 
masses and mixing angles are obtained entirely with the $A_4$ context, 
in good agreement with data.  
In particular, we find a strong correlation between $V_{ub}$ and the 
CP phase $\beta$, thus a decisive  future test of this model can be allowed.

It is one of a powerful guideline to find 
the constraints from neutrinos to grand unification models.  
Discrete symmetries are suitable to decode the flavor problem: 
they can accommodate maximal 2-3 mixing and explain zero 1-3 mixing.  
Moreover they can be the origin of texture zeros or equalities.

\section*{Acknowledgements}

I would like to thank Y.~Koide for his great hospitality and 
organization of this stimulating workshop in Shizuoka.  
I am also grateful to E.~Ma and M.~Tanimoto for their collaboration.


\end{document}